# Do the Young Live in a "Smaller World" Than the Old? Age-Specific Degrees of Separation in Human Communication

Yuxiao Dong*

*Department of Computer Science and Engineering, Interdisciplinary Center for Network Science and Applications (iCeNSA), University of Notre Dame, IN 46556 USA*

Omar Lizardo*

*Department of Sociology, Interdisciplinary Center for Network Science and Applications (iCeNSA), University of Notre Dame, IN 46556 USA and*
*Faculty of Computer Science and Management, Wrocaw University of Technology, 50-370 Wrocaw, Poland*

Nitesh V. Chawla†

*Department of Computer Science and Engineering, Interdisciplinary Center for Network Science and Applications (iCeNSA), University of Notre Dame, IN 46556 USA and*
*Faculty of Computer Science and Management, Wrocaw University of Technology, 50-370 Wrocaw, Poland*

In this paper, we investigate the phenomenon of "age-specific small worlds" using data from a large-scale mobile communication network approximating interaction patterns at societal scale. Rather than asking whether two random individuals are separated by a small number of links, we ask whether individuals in specific age groups live in a small world in relation to individuals from other age groups. Our analysis shows that there is systematic variation in this age-relative small world effect. Young people live in the "smallest world," being separated from other young people and their parents generation via a smaller number of intermediaries than older individuals. The oldest people live in the "least small world," being separated from their same age peers and their younger counterparts by a larger number of intermediaries. Variation in the small world effect is specific to age as a node attribute (being absent in the case of gender) and is consistently observed under several data robustness checks. The discovery of age-specific small worlds is consistent with well-known social mechanisms affecting the way age interacts with network connectivity and the relative prevalence of kin ties and non-kin ties observed in this network. This social pattern has significant implications for our understanding of generation-specific dynamics of information cascades, diffusion phenomena, and the spread of fads and fashions.

The fact that any one individual may be capable of reaching any other one via a relatively short chain of network intermediaries is a surprising property of human social networks[1,2]. This "small world" phenomenon was first documented in a series of classic contact-tracing experiments conducted by Travers and Milgram in the 1960s[3,4], with a recent large-scale Internet-based replication using a cross-nationally diverse population producing results encouragingly close to those of the original study[5]. More recently, with the increasing availability of large-scale network data built from digitally recorded traces of human communication[6–8], the existence of the small-world phenomenon has been successfully established using observational data obtained from large-scale systems featuring millions of actors (nodes) and billions of links[9–11]. One attractive feature of this approach is that it allows for direct calculation of the average number of links separating any two individuals at very close to the whole network level (e.g. the largest connected component in the system). This helps to overcome the key limitation of first generation research on the small world: namely reliance on indirect inference from completed chains obtained from the initial subset of seed nodes. Instead, in large-scale small world research the average of all shortest paths in the network can be calculated directly, although not without computational cost[9,10].

While useful for demonstrating the robust existence of an important property of social networks, a focus on global estimates of the existence of the small world property has to rely on averages taken over all nodes in the network irrespective of node attributes. The disadvantage of this approach is that it may hide structured heterogeneity in the extent to which different node classes are actually well-represented by the average. This becomes more relevant when we consider that people tend to select contacts with similar social characteristics as themselves[12,13], a tendency that is reproduced in the sort of electronic telecommunication platforms that have been the subject of recent attention[14–17]. Because links are not assigned randomly to node-classes, neither are the number of intermediaries separating a given person from others of the same (or different) class. In this respect, in the context of human social networks, it may be more meaningful to investigate the existence of more targeted realizations of the small world phenomenon, especially with respect to node classes defined by socially

_______________

*Y. D. and O. L. contributed equally.
†Email: nchawla@nd.edu

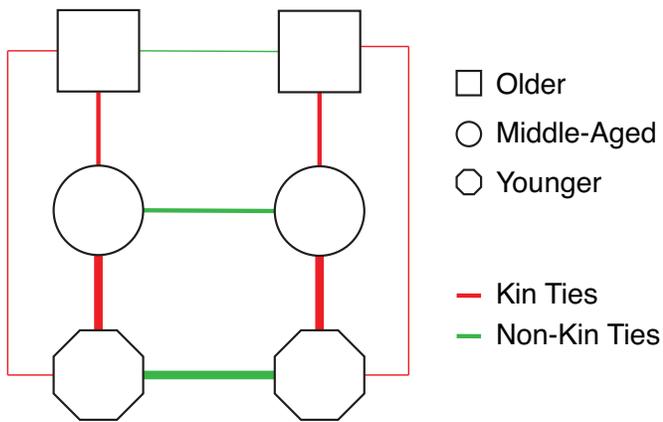

Figure 1: Idealized model of the prevalence and strength of kin and non-kin ties across age groups. Shapes represent three generational groups arranged from younger (octagonal), to middle-age (circle), to older (square). The green edges connecting the shapes represent (idealized) connections among persons who belong to the same age group but who are not biologically related (non-kin ties). The red edges represent (idealized) connections among persons from different age groups who share a biological relation (kin ties). The thickness of the edge indicates the expected relative prevalence and strength (e.g typical communication frequency) for those ties. For the sake of simplicity, cross-generation/non-kin ties are not drawn.

significant attributes such as age, gender, and in some contexts, race and social class.

As a first step in this direction, in this paper we investigate the phenomenon of "age-specific small worlds." Rather than asking whether any two randomly chosen individuals are separated by a small number of intermediaries, we ascertain the extent to which individuals in *the same age group* tend to live in a small world in relation to individuals in the same or other generational clusters. We select age as a focal attribute because it is one of the two (gender being the other one) most powerful traits structuring interaction and sociability in human groups[18–21].

**Age, Social Networks, and the Small World.** What sort of pattern should we expect to observe in terms of the relative strength of the small world phenomenon across age groups? Sociological research on the connection between the age and kin structure, as well as the relationship between non-kin connectivity and life course transitions can be of help in developing some expectations in this regard. Consider the (idealized) model of the connectivity structure between age and kin groups shown in Fig. 1. The figure is meant to encode a series of empirical generalizations taken from relevant work on age, social interaction, and kinship in anthropology and sociology[18,21–25]. The basic idea is that the bulk of informal socializing outside of the family occurs within generations following the principle of age homophily[12]. This means that kin ties are the primary link connecting individuals across generations[22,26]. Kin ties are distinctive because they are largely fixed at birth, are normatively prescribed, and as such display less variation in preva-lence and strength across individuals and groups[18,23]. This has implications for the expected pattern of *cross-generation connectivity* in human societies.

Research in anthropology and sociology points to the historical transformation of the structure of kin ties as societies transition into economic and cultural modernity. As Western (and later non-Western) societies began to industrialize in the the $18^{th}$ and $19^{th}$ centuries, there was a shift towards a "conjugal" (bi-generational) form of family organization[18], and away from tri or quad-generational co-residential living arrangements in which grandparents co-resided with both their children and grandchildren[27]. In this respect, the modal household becomes the bi-generational residence containing only parents and children[28]. In Fig. 1, this is indicated by the thick vertical lines linking the circle (parent) and octagonal (children) generation, and by the relatively thinner vertical links connecting the circle and square (grandparent) and the even thinner lines connecting the octagonal and square.

In addition, note the declining strength of within-generation, cross-kin connectivity as we move up from the youngest to the older groups in Fig. 1. This encodes a series of stylized facts from sociological work on the relationship between age and social networks, having implications for the expected pattern of *within-generation connectivity* at the societal level. First, with regards to younger people, sociological work on the subject shows that, free from the demands of work, childcare, and other mid-life responsibilities, younger people are better able to devote relatively large amounts of time to within-generation socializing outside the family, increasing their connectivity within this age stratum[29]. In addition, younger individuals tend to spend the majority of their time inhabiting social institutions (such as schools) that encourage same-generation non-kin peer group formation and promote sociable interaction[30]. Second, middle-aged individuals, while continuing to have active dispositions and capacities for socializing with same-age non-kin others, experience a variety of life events that lead to a decline in connectivity. These include transition into marriage, full-time employment, and parenthood[31,32]. Finally, a long line of research in sociology, anthropology and gerontology demonstrates that older persons experience strong declining attachments to same age peers, with all indicators of sociability experiencing steep drops. These include non-kin contact volume, emotional closeness, and time spent in the presence of others[19,21,25,33,34,34]. This also means that as individuals age and lose same-generation non-kin ties, cross-generational connections to children and other relatives come to form a larger proportion of their remaining network[22,26].

**Implications for Age-Specific Small Worlds.** Because the small world property is premised on the relative connectivity of individuals[35] in relation to others, the existence of combined age and kin effects on social interaction volume should result in predictable consequences for the relative extent to which individuals of different age groups live in a small world. Generally, the less connected the members of a given age group are to others of a given node class (e.g. same or different generation), the less likely they are to be able to reach those others via a small number of intermediaries. Given the empirical patterns encoded in Fig. 1, we should then expect that: (*a*) younger individuals should live in the smallest of worlds, especially with respect to same-generation others. In addition, (*b*) given the existence of relatively strong ties to parental generation (via the bi-generational household residence mechanism), they should also be separated by a relatively small (but larger than the same-generation quantity) number of intermediaries from members of the parental generation (and vice versa). However, (*c*) relatively fractured attachments to the grandparent's generation produced by the same bi-generational household structure, should put young people at a longer sociometric distance from their most older counterparts (and vice versa), while (*d*) middle-aged individuals should be in the next "least small" world tier with respect to same-generation peers. That is, their separation from same-generation others should be larger than that of corresponding to their children. Middle-aged individuals should also (*e*) be relatively close to members of the parental generation via intermediary kin ties. Finally, (*f*) older individuals should live in the "least small" world with respect to same-generation peers, as ties to same-generation others are selectively pruned leaving only kin-tie mediated attachment to middle-aged members of their sons and daughters generation as their primary source of sociability.

## Results

We begin by addressing the question of whether we can identify age-specific small worlds. To do so, we use a large-scale mobile phone data capturing patterns of communication at a societal scale. The data is comprised of more than one billion voice calls and short messaging records spanning two consecutive months—August and September—in the year 2008 representing about one fifth of the population of a large industrialized country. These data are appropriate for our research goals as they have been used profitably in previous studies establishing strong regularities in human communication and mobility behavior[8,36,37]. To represent this large-scale communication system as a network, we place an edge between two users if and only if they have reciprocal communications (voice calls or text messages)[8] within the observation time-frame, ensuring that the links capture significant social interactions and relationships.

It is possible that any conclusions regarding age-effects in small-world behavior might be systematically affected by the two month observation window or may not be unique to the generation-specific processes that we outlined earlier. We address these issues in four ways. First, we construct communication networks of increasing temporal scale (using a one-week window), and examine whether our results hold within each cumulative time slice. Second, we also examine whether the values of key quantities, such as the average shortest path lengths, show signs of convergence as we extend the temporal window. Observing such saturation behavior would indicate that two-months are sufficient to extract steady-state properties of the system. Third, we examine whether there are differences in small world behavior across age-by-gender groups, given that gender is a distinct, but equally significant, node-level trait affecting connectivity patterns. Null findings in this respect would provide additional evidence for the generational mechanisms proposed. Finally, we trace patterns of cross-generation connectivity across age-levels and examine whether they provide evidence for the connectivity mechanisms illustrated the idealized model depicted in Fig. 1.

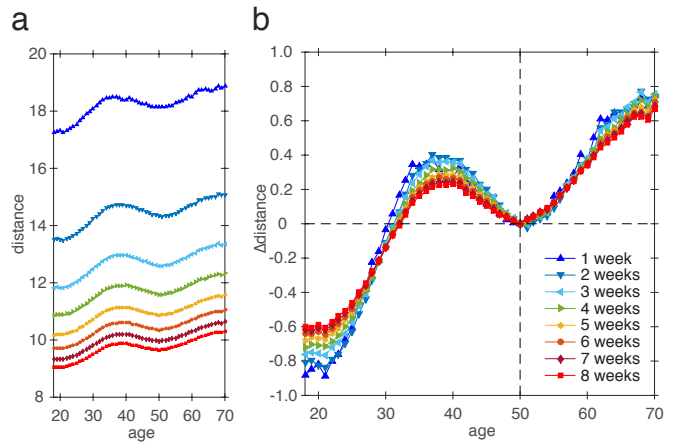

Figure 2: Age-specific small worlds across different time-frames in the mobile network. The average degrees of separation vary as a function of age (*a*); The relative variations of age-specific degrees of separation is constant (*b*), that is, in each time-frame the average distance of the 50-year-old people is scaled to 0. See Supplementary Figures 1 and 2 for the results in phone call and text messaging networks.

**The Young Live in A Smaller World.** The results of the age-specific small world analysis are shown in Fig. 2a. We conduct all analyses on the mobile (phone calls and text messages), phone call (CALL), text messaging (SMS) networks, as the results are the same regardless of what communication channel we use as a connectivity criterion (see Supplementary Figures 1 and 2). The basic empirical patterns are consistent with expectations. The average shortest path connectivity in the mobile communication network increases steadily with age, until about age 35; it then declines until about age 50 and then rises steadily again into old age. Note that the age markers for the period of increasing "small worldness" for adults (35-50) correspond closely to the ages at which members of advanced industrial societies

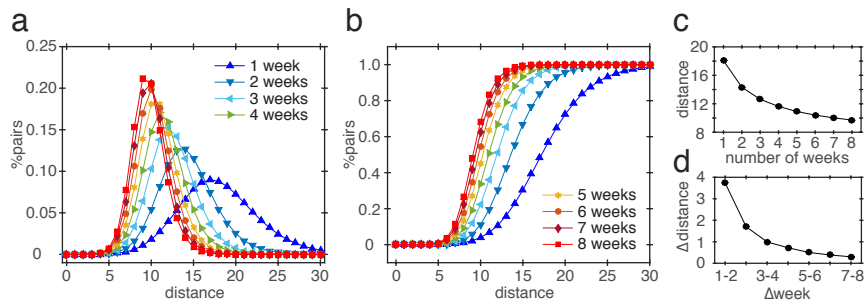

Figure 3: **Convergence in shortest path estimates with increasing temporal window.** The probability mass functions (PMF) of shortest path lengths (distances) across different number of weeks (a); The cumulative distribution functions (CDF) of distances across different number of weeks (b); The average distance between each pair of users and time (c); The gap between the distances of two consecutive weeks (d). For example, when $x$ is 3-4, the corresponding $y$ value represents the difference between the average shortest path lengths observed from the 3-week network and 4-week network. See Supplementary Figures 3 and 4 and Supplementary Note 1 for the results in phone call and text messaging networks.

will be forming "downward" kin ties to their children.

Fig. 2b shows that age specific average-shortest path distances exhibit the same relative trend with respect to age regardless of the time-window used. To construct the plot, we took the estimates of the average shortest path connectivity of the 50-year-old population as our reference point in each network, putting all time-slice-specific trend-lines in the same scale. The plot shows that the relative small-world gaps between members of different generations is essentially identical across time-windows and can already be observed in the most restricted (one-week) version of the data. These result are consistent with the claim that two-months of communication data is sufficient to establish large-scale regularities with respect to shortest-path behavior in this network. Fig. 3 provides corroborating evidence for this claim, showing that the accuracy gains of adding additional layers of data decrease dramatically after we cross the three-week cumulative time-slice, with estimates of the network property under investigation (average shortest path) converging around a similar steady-state value after the six-week mark (see Supplementary Figures 3 and 4 for the results in phone call and text messaging networks).

**The Young are Close to the Young and the Old are Far from the Old.** What are the sources of the small world advantage of young people? To answer this question, we compute average shortest path distances across dyad classes composed of people of different age groups (ranging from 18 through 70). This is shown in Fig. 4a. As expected, the small world advantage of younger individuals comes from their relative closeness to their same age counterparts (blue shaded area in the lower left-hand corner of each subplot) coupled with their relative closeness to individuals in their parent's generation (about 20 to 30 years older). This is consistent with sociological work suggesting that the first pattern is due to the formation of non-kin same generation ties (although ties to siblings in the same generation are also included here), while the latter are due primarily to kin ties to parents (and indirectly to other members of the parental generation). Individuals between the ages of 35 and 50 end up being sociometrically closer to their younger counterparts (offspring generation) than they are to their own generation, thus explaining the relative decline in average shortest path distances for individuals within this age range. This result is consistent with sociological research pointing to the disruption of same-generation non-kin ties with middle-aged life transitions, and the relative stability and durability of kin ties to offspring given their non-elective status[18,32].

As shown in the red-shaded area in the upper-right hand corner of the plot, the reason why older individuals live in the "least small world." is due to their relatively large sociometric distance from members of the same-generation and that of their immediately preceding (offspring) age group. This is consistent with work showing steady decline in sociability and connectivity with in elective (non-kin) ties leaving older persons with non-elective (kin) ties as their only source of connectivity[22]. As with our previous results, relative age-based patterns of cross-generational connectivity observed in the 8-week network are also consistently observed in the 1- to 7-week networks (Figs. 4d–j). This robustness check shows that the age-specific small world effect is independent on restrictions on the temporal window covered by our data.

Fig. 4b shows a heatmap illustrating what happens when we shuffle the demographic attributes of each vertex in the network (leaving both the network structure and the proportion of vertices belonging to a given age group intact) while computing the average shortest path distances across age groups for fifty different realizations of the reshuffled network (see Materials and Methods for details). As shown by the homogeneous coloring across the figure, age-group differences in average shortest-path distances to members of other age groups disappear, and all age groups converge to the average geodesic distance for all pairs in the mobile network ($\mathcal{L} \approx 9.7$). This suggests that, consistent with our account, differences across age groups in "small worldness" emerge as a result of systematic preferences and constraints generating specific within and cross-generation social attachments in human populations[12].

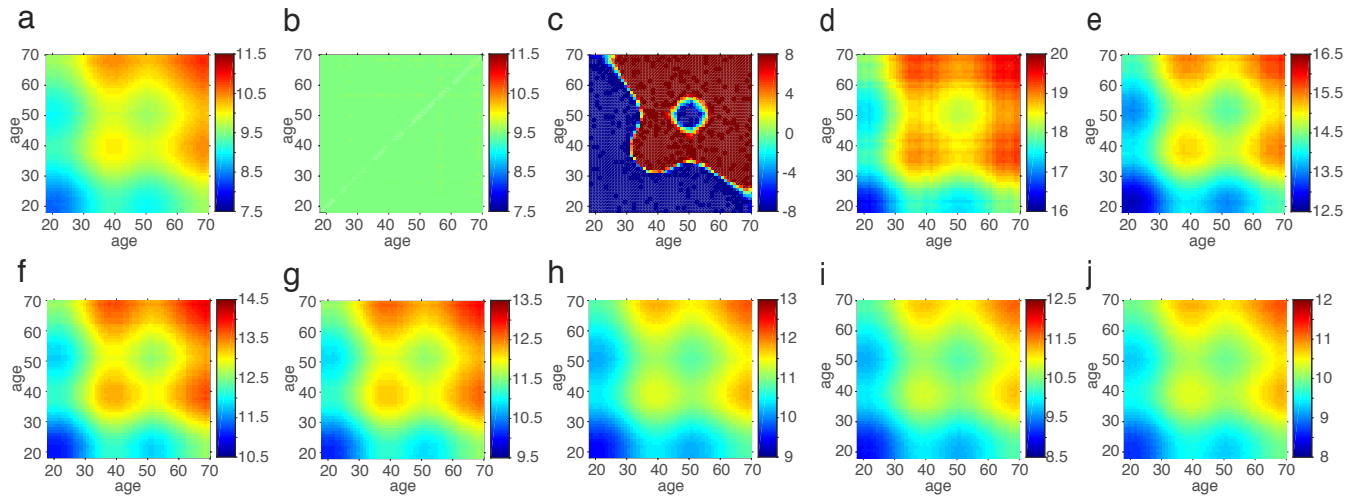

Figure 4: **Average degrees of separation across age groups.** The spectrum color represents the average shortest path length in the 8-week Mobile network (*a*), shuffled average shortest path length (*b*), and $\mathcal{Z}$-score value (*c*) between two people of age indicated by *x*- and *y*-axes. The spectrum color in figures *d, e, f, g, h, i*, and *j* represents the average shortest path length in the 1-, 2-, 3-, 4-, 5-, 6-, and 7-week mobile networks. See Supplementary Figures 5 and 6 for the results in phone call and text messaging networks.

Fig. 4c shows a heatmap of the distribution of $\mathcal{Z}$-scores obtained from comparing the observed average geodesic distances across age groups against what we would have expected by chance (from the fifty reshuffled realizations of the network as given in Fig. 4b). The results confirm that younger individuals live in smaller worlds in relation to same generation peers and older generation contacts than we would expect by chance, while older individuals live in larger than expected small world in relation to same generation peers and members of the immediately preceding generation (middle-aged individuals).

**Null Gender Differences in Age-Specific Small Worlds.** To provide corroborating evidence that the mechanisms generating differences in small world behavior are unique to generational node classes, we investigate whether there are gender effects in age-specific small worlds. Looking at differences between men and women is relevant, since other than age, gender is the one characteristic that has been shown to systematically impact aspects of communication behavior in social networks[15,31]. However, if the model presented in Fig. 1 is on the right track we should find little or no evidence of gender-by-age specificity in small worlds, since sociological work shows that the mechanisms generating the age-specific small-world behavior are common to both men and women.

As Fig. 5 shows, we find that the pattern of decreasing "small-worldness" as persons age is common to men and women (see Supplementary Figures 7 and 8 for the results in phone call and text messaging networks). The one exception is the slightly stronger increase in "small worldness" for women between the ages of 30–50 in relation to men of the same age. This pattern of results is consistent with the downward (offspring generation) kin-based connectivity mechanism proposed to explain this effect, as mothers are more likely to maintain regular interactions with their children than fathers. Notably, we find that average shortest path estimates do not differ across dyad pairs classified according to the gender mix (Figs. 5b–c). Our analyses show that the shortest path connectivity between two females (F-F), one male and one female (M-F), and two males (M-M) follows highly overlapping distributions (Fig. 5b) and are not sensitive to time-window restrictions in the data (Fig. 5c). Figs. 5d-f show that the relative sociometric distance between nodes based on age-classes does not depend on the gender mix as the same pattern of the young being close to the young being close to their same age peers and the old being far from other old people is replicated for same gender (d, f) and different gender (e) dyad classes. Given that previous studies have shown that cross-gender interactions are consistently more intense and frequent than those between same-gender pairs in different communication channels[9,17], our findings suggests that relative small world differences between age groups are not generated by heterogeneity in the characteristic link strengths across age-classes.

**Evidence for Proposed Connectivity Mechanisms.** As we noted earlier, sociological and anthropological work on age and social networks suggests that the mechanism generating age-specific small worlds is that the cross-age-group connectivity distribution is systematically different for older and younger persons. More specifically, same-generation (primarily non-kin) sociability should steadily decline and be replaced by in-creasing cross-generation (primarily kin-based) sociability. To examine whether we can observe evidence of this mechanisms in this network, Fig. 6a plots the proportion of ages for each age group that link them to same generation (plus or minus five years difference), older generation (between 20 to 30 years older) and younger generation (between 20 to 30 years younger) groups. All rates are calculated from the mobile electronic communication network.

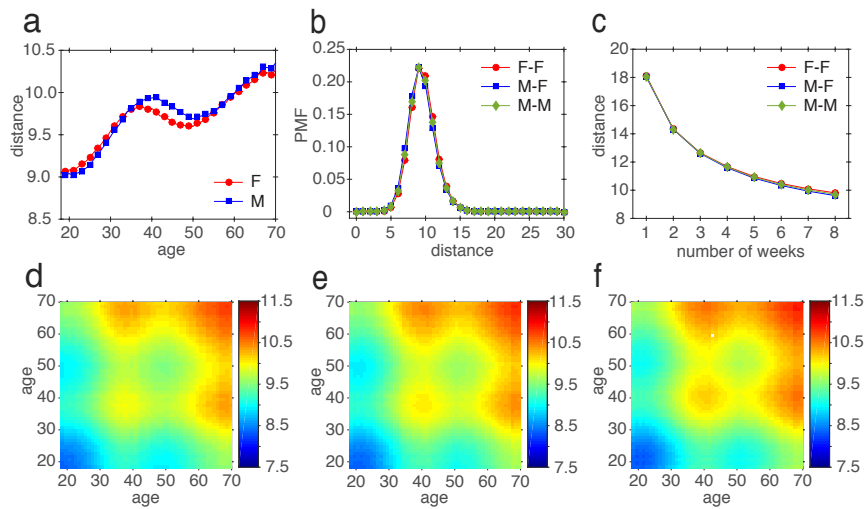

Figure 5: **Gender-specific small worlds across age groups.** The average distances by age do not vary a lot for female (F) and male (M) in the 8-week mobile network ($a$); the probability mass functions of shortest path distances between three different gender pairs overlap with each other in the 8-week network ($b$); The average distances between different gender pairs are the same in all eight networks of different length of time-frames ($c$); the spectrum color represents the average shortest path lengths between two females ($d$), one male and one female ($e$), and two males ($f$) in the 8-week mobile network. The strong similarities among the three heatmaps suggest relative age-specificity of mobile small worlds does not depend on gender in a strong way. See Supplementary Figures 7 and 8 for the results in phone call and text messaging networks.

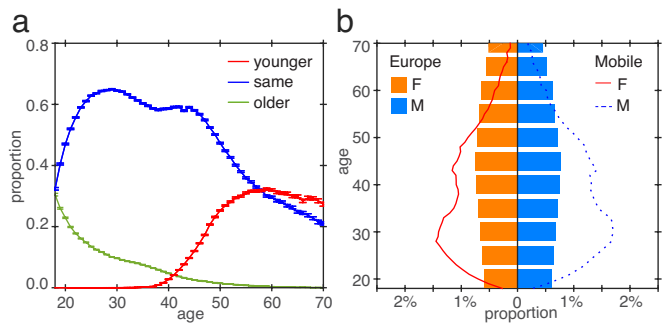

Figure 6: **Connectivity mechanisms behind age-specific small worlds.** The proportion of one's contacts of different age groups conditioned as a function of the person's own age ($a$). Specifically, one's contacts of the "same" generation are denoted as those aged between $x$-5 and $x$+5, where $x$ represents his or her age, the "older" generation aged between $x$+20 and $x$+30, and the younger generation aged between $x$-30 and $x$-20 (The mean values are observed at a 95% confidence interval); The population distribution observed from the mobile data is different from the European population distribution at the same year, that is, 2008. See Supplementary Figures 9 and 10 for the results in phone call and text messaging networks.

The findings provide strong evidence for the idealized pattern depicted in Fig. 1, suggesting that these are the mechanisms behind the age-specific small world effects that we observe. Younger individuals (e.g. between the ages of 20 and 35) have relatively high rates of communicative interaction with both their same age peers and those in the immediately preceding (parental) generation. However, as we move up along the $x$-axis, we see a steady decline in same-generation sociability and its gradual replacement by cross-generation sociability (20 to 30 years younger). This is indicative of attrition in same-generation ties for older individuals and their replacement with cross-generation ties to the immediate kin (child) generation. The two lines cross at about 60 years of age, which is close to the institutionally mandated time for retirement from work activity in industrialized Western societies (such as the one from which the mobile network originated), providing further support for the model.

## Discussion

Previous work has shown the "small world" property to be a counter-intuitive but robust signature of human social networks. The classic work by Milgram[3] as well as more recent replications using email chains[5] used experimental strategies aimed at inferring average network diameter from the average length of completed chains. More recent work using social media and electronic communication data allows for the computation of average shortest paths at a societal or even "planetary" scale[9–11]. However, most of this work remains focused on the small world property as a feature of the entire network, but has not looked at vertex-level heterogeneity in the exis-tence of this property.

In this paper, we ask the question of whether there are age-specific small worlds. Using a large-scale mobile communication network that approximates the volume of communication of a large-scale industrialized society[8,38], we ask not whether any two random individuals are separated by a relatively small number of intermediaries. Rather, we ask whether individuals in different age groups live in more or less small worlds in relation to members of other groups. This question is important since age is one of the few characteristics that has been shown to structure human interaction in all human societies[39].

Our results reveal systematic heterogeneity in the extent to which people of different ages can be said to live in "more or less" small worlds. The pattern of this heterogeneity is, in its turn, predictable from well-known regularities uncovered in anthropology and sociology related to the relationship between age, sociability, changing structure of generational living arrangements in modern societies, and the relative rates of kin-based and non-kin-based connectivity throughout the life-course (see Fig. 1). Younger individuals live in the smallest of worlds, both in relation to same-age peers and cross-generation consociates, while older individuals live in the "least small" of worlds, being particularly likely to be separated by a larger number of intermediaries from same-generation peers.

These results have important implications because the small world property of human social networks lies behind a variety of phenomena associated with processes of cultural transmission, the emergence of information cascades, and other diffusion processes. As a rule, shortest connectivity paths between persons facilitate the fast spread of information and thus contribute to the large-scale adoption of novel beliefs, behaviors, practices, and products[2,40]. Our results thus imply that in any given modern society, due to their greater sociometric proximity to both same-generation and cross-generation others, younger persons are more likely to serve as the most effective seeds and most likely conduits for the rapid spread of novel information, behaviors, practices, and any other element that may be subject to "contagion" and diffusion dynamics. Older individuals, on the other hand, due to their greater sociometric isolation from other age groups, are the least likely to play this role. This implication is consistent with work in sociology and marketing showing that such phenomena as fads, fashions, and information/behavior cascades occur more frequently among the young[41], and that members of older generations are generally dependent on younger individuals to keep abreast of novel behaviors, products, and activities[42]. Our work thus reveals that these long standing observations have an intuitive basis in the sociometric location of the young in relation to the old.

Our results also imply that greater sociometric isolation of older individuals will result in their being the last to hear or be exposed to novel "viral" practices, beliefs, and objects net of any dispositional "conservatism" that may come with advanced age[43]. Thus even older individuals who may be potentially open to new experiences and be likely candidates for the adoption of innovations, will be at a structural disadvantage. However, our argument and results do suggest that *if* older persons do experience exposure it is more likely to come from cross-generational next of kin ties (most likely children) than from non-kin same-generation peers. In this respect, the existence of various "generation gaps" in attitudes, behaviors, and practices may be as much of a product of the qualitatively distinct social structural position of the young and the old as it is of cohort-based, period-based, or aging-dynamics.

In this paper, we have provided a model and a set of tools for how to investigate heterogeneity in "generic" properties of large-scale networks across vertex attributes. Future work can build on our current effort and examine the extent to which heterogeneity in the small world (and other well-defined network properties) that have been primarily investigated irrespective of the categorical attributes of vertices in human social networks do vary in a structured way according to those attributes, while outlining the implications of this variation for our understanding of important structural and dynamics processes in such networks.

## Materials and Methods

**Mobile Phone Networks.** We use a mobile phone data comprised of more than 1 billion voice call and short messaging records spanning in two consecutive months—August and September—in 2008[17,36]. To represent the human communication behavior in networks, we place an edge between two users if and only if they have reciprocal communications (voice calls or text messages) within the observation time-frame[8]. To investigate the evolution of mobile small worlds, we choose to use one week as the time unit to create networks of different durations (weeks). Specifically, we use the first week (Monday, August $4^{th}$ to Sunday, August $10^{th}$, 2008) of communication logs to construct the 1-week network. Similarly, the $k$-week network ($1 < k \leq 8$) was extracted from the first $k$ consecutive weeks of data, meaning that the first day of each network always starts from Monday, August $4^{th}$, 2008 and the last day of the 8-week network ends at Sunday, September $28^{th}$, 2008. Further we extract the giant component as the experimental network from each network[8–10]. In this way we construct eight mobile networks of different length of durations from the communication logs, with the largest and longest-spanning network, the 8-week one composing of 5,171,066 nodes and 9,885,493 undirected edges. The order and size of the eight mobile networks are listed in Table 1. Our visualization shows that the combined mobile networks obey the densification power law[44] with a good fit, that is, the number of edges grow superlinearly in the number of nodes in mobile communication networks (see Supplementary Figure 11).

Table 1: The statistics of eight mobile networks.

| Networks | 1-week | 2-week | 3-week | 4-week | 5-week | 6-week | 7-week | 8-week |
|---|---|---|---|---|---|---|---|---|
| #nodes | 1,406,743 | 2,698,575 | 3,444,931 | 3,958,354 | 4,371,045 | 4,686,770 | 4,948,254 | 5,171,066 |
| #edges | 1,672,693 | 3,659,144 | 5,119,451 | 6,314,822 | 7,402,307 | 8,336,223 | 9,153,808 | 9,885,493 |

In the resulting 8-week mobile network, 89% of nodes are associated with the corresponding users' gender and age information. We calculate the shortest paths between all pairs of users and report the results between those with known gender and age attributes. The demographic distributions of users in the 8-week mobile network and the European population are presented in Fig. 6b. Overall there are 45% of female users and 55% of male users, and both female and male users between the ages of 18 and 55 are overrepresented in mobile communication compared to the true population, while teenagers and seniors on the other hand are underrepresented.

**Shortest Paths in Big Networks.** In this work, rather than employing the sampling and probabilistic methods used in previous work[9,10], we instead leverage a 48-core CPU computing server to determine the *exact* shortest path length between *all* pairs of nodes, that is $s = n \times (n-1)/2$ pairs, where $n$ is the number of users in each network. We use the parallel breadth-first search algorithm[1] to compute the shortest paths between all pairs of users, and more essentially, during each step of search, to record the length of the shortest path between two users specified by their gender and age information. In the parallel algorithm, $n/48$ nodes' distances to all $n$ nodes are allocated to one CPU for computation. For example, to compute the shortest path distances between $s \approx 1.33 \times 10^{13} (n = 5,171,066)$ pairs of users in the largest mobile network (8-week), each CPU is responsible for $s/48 \approx 2.8 \times 10^{11}$ pairs of users. By using the computer server with Quad 12 core 2.3 GHz Intel Xeon CPUs E7-4850 (48 cores in total), we are able to compute the exact shortest path length between all pairs of users within 37 hours for the 8-week mobile network.

**Null Model.** We validate the statistical significance of the differences on shortest path lengths between users of different age and gender groups. The idea of the statistical test is to compare the gender- and age-based shortest path length $x$ from the data to those $\{\tilde{x}\}$ provided by a null model, wherein the users' gender and age are randomly shuffled. In this article, we leverage a classical null model that was used in attributed networks[15]. On the null model, we first randomly assign users' gender and age on the underlying structure of the communication networks, and then compute the lengths of the shortest path between pairs of users of randomly allocated gender and age. The random assignment of users' gender and age and following computation of shortest paths in the network are simulated for fifty times. Accordingly, we calculate the mean $\mu(\tilde{x})$ and standard deviation $\sigma(\tilde{x})$ of the shortest path lengths $\{\tilde{x}\}$ between users with shuffled gender and age by the null model. We use $\mathcal{Z}$-score to examine the gap between the real data $x$ and the randomly shuffled results $\{\tilde{x}\}$ generated by the null model, wherein $\mathcal{Z}(x) = \frac{x - \mu(\tilde{x})}{\sigma(\tilde{x})}$. A $\mathcal{Z}$-score of 0 indicates that there exists no difference between the real data and the null model. A positive (negative) $\mathcal{Z}$-score represents that the empirical data is above (below) the null model result. $|\mathcal{Z}(x)| \geq 3.3$ (corresponding to $p$-value $\leq 0.001$) represents that the observation from the empirical data is extremely statistically significant.

**Acknowledgments**
We sincerely thank Dr. Albert-László Barabási for sharing the mobile phone data. This work is supported by the Army Research Laboratory under Cooperative Agreement Number W911NF-09-2-0053 and the National Science Foundation (NSF) Grants BCS-1229450 and IIS-1447795.

**Author contributions**
Y.D., O.L. and N.V.C. designed the research; Y.D., O.L. and N.V.C. performed research; Y.D., O.L. and N.V.C. analyzed data and results; Y.D., O.L. and N.V.C. wrote the paper. Y.D. and O.L. contributed equally to this work.

**Additional information**
The authors declare no conflict of interest.

---

[1] The code will be made publicly available.

# Supplementary Information

## Supplementary Figures

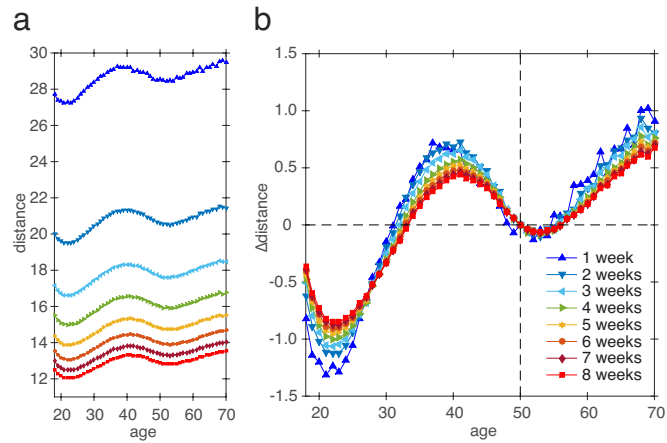

Supplementary Figure 1: Age-specific small worlds across different time-frames in the CALL network. The average degrees of separation vary as a function of age ($a$); The relative variations of age-specific degrees of separation is constant ($b$), that is, in each time-frame the average distance of the 50-year-old people is scaled to 0.

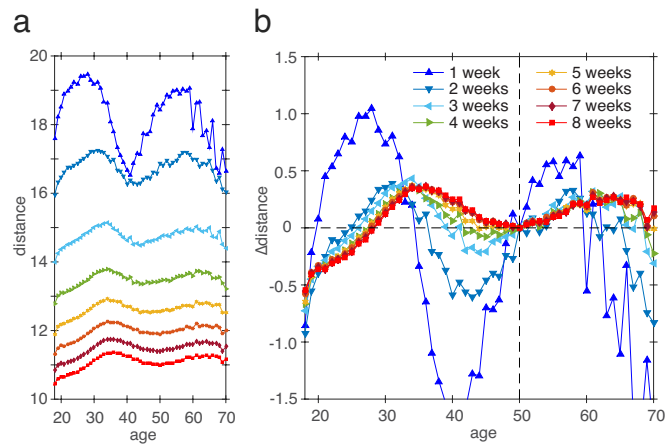

Supplementary Figure 2: Age-specific small worlds across different time-frames in the SMS network. The average degrees of separation vary as a function of age ($a$); The relative variations of age-specific degrees of separation is constant ($b$), that is, in each time-frame the average distance of the 50-year-old people is scaled to 0.

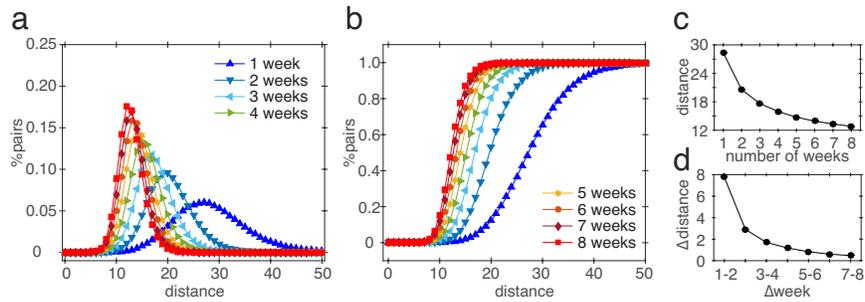

Supplementary Figure 3: Convergence in shortest path estimates with increasing temporal window in the CALL network. The probability mass functions (PMF) of shortest path lengths (distances) across different number of weeks (a); The cumulative distribution functions (CDF) of distances across different number of weeks (b); The average distance between each pair of users and time (c); The gap between the distances of two consecutive weeks (d). For example, when $x$ is $4 - 3$, the corresponding $y$ value represents the difference between the average shortest path lengths observed from the 3-week network and 4-week network.

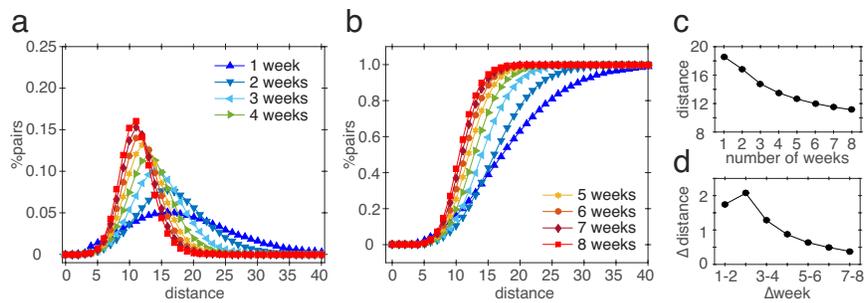

Supplementary Figure 4: Convergence in shortest path estimates with increasing temporal window in the SMS network. The probability mass functions (PMF) of shortest path lengths (distances) across different number of weeks (a); The cumulative distribution functions (CDF) of distances across different number of weeks (b); The average distance between each pair of users and time (c); The gap between the distances of two consecutive weeks (d). For example, when $x$ is $4 - 3$, the corresponding $y$ value represents the difference between the average shortest path lengths observed from the 3-week network and 4-week network.

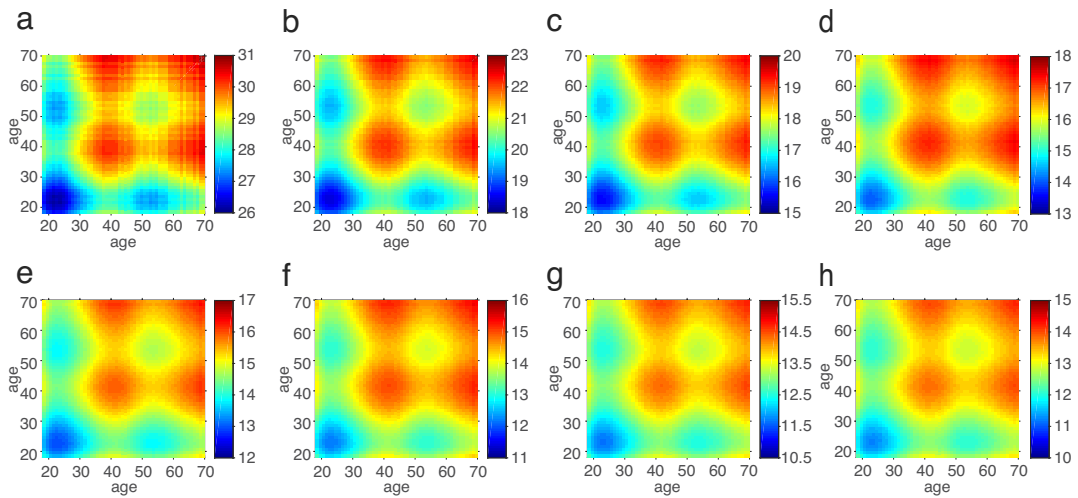

Supplementary Figure 5: Average degrees of separation across age groups in the CALL network. The spectrum color represents the average shortest path length in the 8-week Mobile network ($a$), shuffled average shortest path length ($b$), and $\mathcal{Z}$-score value ($c$) between two people of age indicated by $x$- and $y$- axes. The spectrum color in figures $d$, $e$, $f$, $g$, $h$, $i$, and $j$ represents the average shortest path length in the 1-, 2-, 3-, 4-, 5-, 6-, and 7-week Mobile networks.

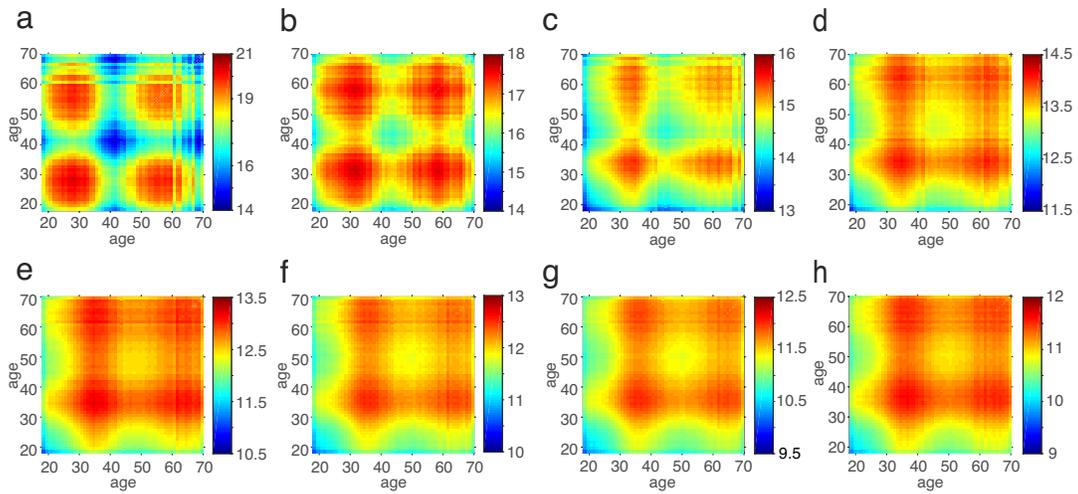

Supplementary Figure 6: Average degrees of separation across age groups in the SMS network. The spectrum color represents the average shortest path length in the 8-week Mobile network ($a$), shuffled average shortest path length ($b$), and $\mathcal{Z}$-score value ($c$) between two people of age indicated by $x$- and $y$- axes. The spectrum color in figures $d$, $e$, $f$, $g$, $h$, $i$, and $j$ represents the average shortest path length in the 1-, 2-, 3-, 4-, 5-, 6-, and 7-week Mobile networks.

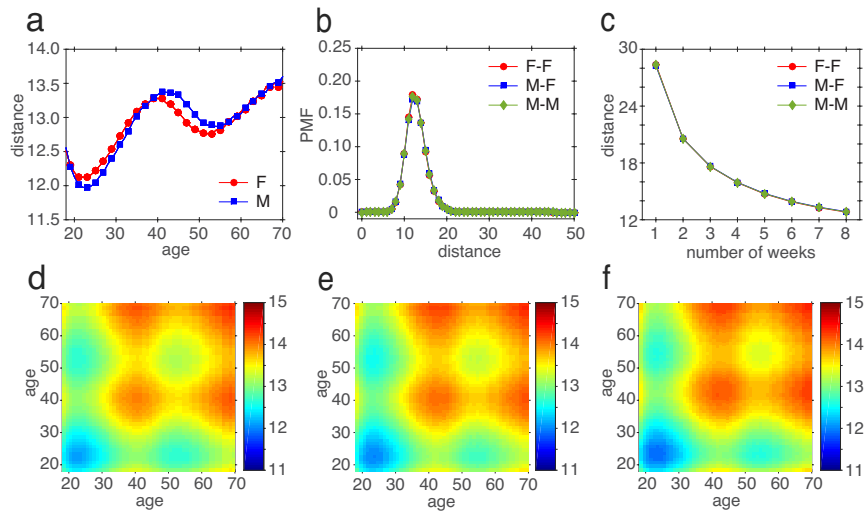

Supplementary Figure 7: Gender-specific small worlds across age groups in the CALL network. The average distances by age do not vary a lot for female (F) and male (M) in the 8-week mobile network ($a$); the probability mass functions of shortest path distances between three different gender pairs overlap with each other in the 8-week network ($b$); The average distances between different gender pairs are the same in all eight networks of different length of time-frames ($c$); the spectrum color represents the average shortest path lengths between two females ($d$), one male and one female ($e$), and two males ($f$) in the 8-week mobile network. The strong similarities among the three heatmaps suggest relative age-specificity of mobile small worlds does not depend on gender in a strong way.

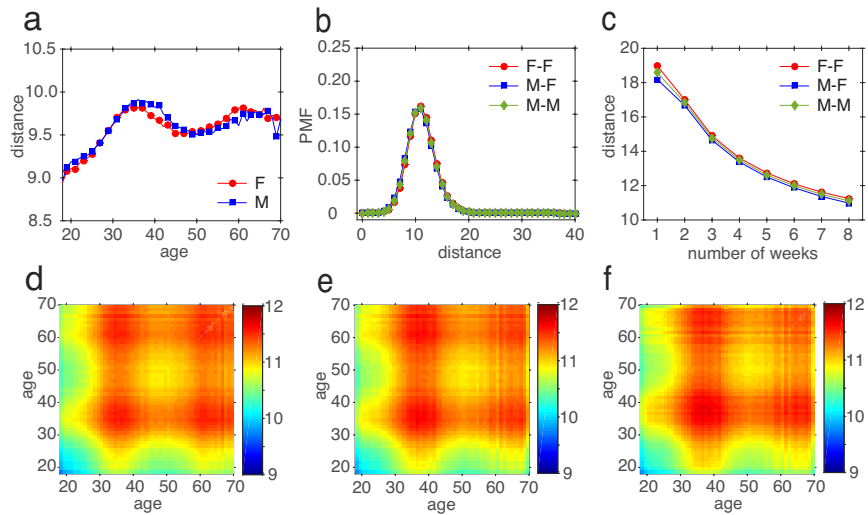

Supplementary Figure 8: Gender-specific small worlds across age groups in the SMS network. The average distances by age do not vary a lot for female (F) and male (M) in the 8-week mobile network ($a$); the probability mass functions of shortest path distances between three different gender pairs overlap with each other in the 8-week network ($b$); The average distances between different gender pairs are the same in all eight networks of different length of time-frames ($c$); the spectrum color represents the average shortest path lengths between two females ($d$), one male and one female ($e$), and two males ($f$) in the 8-week mobile network. The strong similarities among the three heatmaps suggest relative age-specificity of mobile small worlds does not depend on gender in a strong way.

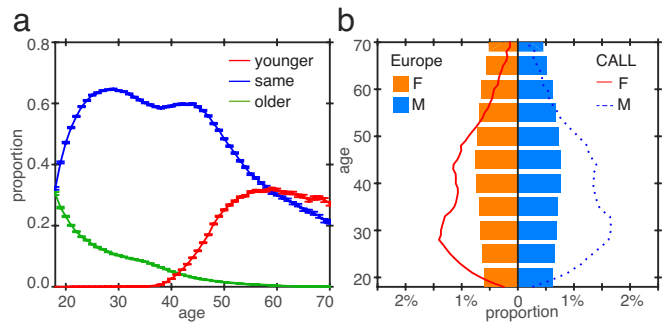

Supplementary Figure 9: Connectivity mechanisms behind age-specific small worlds in the CALL network. The proportion of one's contacts of different age groups conditioned as a function of the person's own age ($a$). Specifically, one's contacts of the "same" generation is denoted as those aged between $x$-5 and $x$+5, where $x$ represents his or her age, the "older" generation aged between $x$+20 and $x$+30, and the younger generation aged between $x$-30 and $x$-20 (The mean values are observed at a 95% confidence interval); The population distribution observed from the mobile data is different from the European population distribution at the same year, that is, 2008 ($b$).

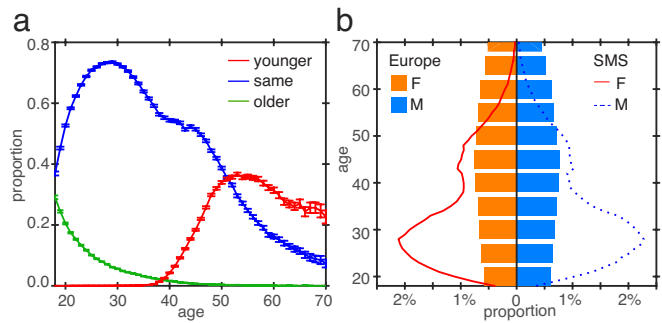

Supplementary Figure 10: Connectivity mechanisms behind age-specific small worlds in the SMS network. The proportion of one's contacts of different age groups conditioned as a function of the person's own age ($a$). Specifically, one's contacts of the "same" generation is denoted as those aged between $x$-5 and $x$+5, where $x$ represents his or her age, the "older" generation aged between $x$+20 and $x$+30, and the younger generation aged between $x$-30 and $x$-20 (The mean values are observed at a 95% confidence interval); The population distribution observed from the text messaging data is different from the European population distribution at the same year, that is, 2008 ($b$).

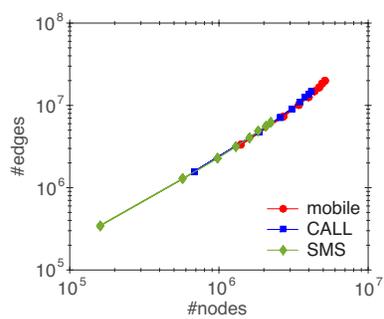

Supplementary Figure 11: The size of mobile phone networks as a function of their orders in log-log scales. Three networks extracted from different channels obey the densification power law[44] with a close slopes.

# Supplementary Tables

Supplementary Table 1: The statistics of eight phone call networks (CALL).

| Networks | 1-week | 2-week | 3-week | 4-week | 5-week | 6-week | 7-week | 8-week |
|---|---|---|---|---|---|---|---|---|
| #nodes | 683,422 | 1,856,733 | 2,587,069 | 3,087,363 | 3,479,397 | 3,771,458 | 3,996,406 | 4,176,011 |
| #edges | 786,952 | 2,385,335 | 3,606,013 | 4,600,727 | 5,499,004 | 6,266,175 | 6,915,723 | 7,482,933 |

Supplementary Table 2: The statistics of eight text messaging networks (SMS).

| Networks | 1-week | 2-week | 3-week | 4-week | 5-week | 6-week | 7-week | 8-week |
|---|---|---|---|---|---|---|---|---|
| #nodes | 159,745 | 570,219 | 972,996 | 1,305,151 | 1,596,555 | 1,838,026 | 2,052,003 | 2,241,307 |
| #edges | 172,657 | 639,635 | 1,141,875 | 1,596,942 | 2,026,486 | 2,411,111 | 2,770,033 | 3,101,637 |

## Supplementary Notes

**Supplementary Note 1: Mobile phone call and text messaging networks**

We examine the age-specific small worlds in the phone call (CALL) and text messaging (SMS) channels separately. We observe that the age-specific small worlds in SMS are in several ways different from those in the CALL network (Supplementary Figs. 1, and 1), which tells a similar story with the combined mobile network. First, the most noticeable difference lies in the relatively flat trend of average distance across age groups in the SMS network. That is, the contrast of the scales of small worlds in different age groups is not as notable as the case in phone call. Nevertheless, the argument that the young still live in the smallest world still holds in this channel. Due to the limited size of text messaging users of age over 60, however, it is difficult to draw the same conclusion that the older live in the largest small world in the SMS network. Second, the average distances across different age groups in SMS are consistently shorter than the numbers in the CALL network, suggesting a smaller messaging world. Third, the young aged 18, 19, and 20 are connected via slightly longer chains of intermediaries through phone calls than the young of ages 21 to 25, while the phenomenon is not observed from the SMS network. Similarly, the middle-age peak value in the CALL network appears in ∼40 years old, while this peak is reached at age 35 in the SMS network. In addition, the middle-age valley in the SMS network occurs about five years early than the CALL network. These indicate that there exists a roughly five-year shift in age-specific small worlds from the CALL network to SMS. This finding can be further evidenced from Supplementary Figs. 5 and 6, where the spectrum in Supplementary Fig. 6h can be viewed as the upper-right corner to bottom-left corner shift of Supplementary Fig. 5h.

Accordingly, we examine the proportion of one's contracts of different ages in the CALL and SMS networks, respectively (see Supplementary Figs. 9a and 10a). The distributions in these two specific channels are coupled with that in the combined mobile network (Fig. 6), indicating the same mechanisms in mobile communication that drive the formation of the smallest world among the young and the least small world among the older people. Similar to the combined mobile phone usage, phone call and text messaging populations follow similar distributions (see Supplementary Figs. 9b and 10b), including the over-representation of people aged between 18 and 55 years old, as well as the under-representation of age 55+ people compared to the European population at the same time—2008. In general, the phone call and text messaging systems exhibit similar patterns in age-specific small worlds, suggesting the robustness of our discoveries in the combined mobile channel.